\documentclass{article}
\usepackage{hyperref}
\usepackage{algorithm}
\usepackage{algpseudocode}
\usepackage{graphicx}
\usepackage{multirow}
\usepackage{amssymb}

\newcommand{\ie}{i.e.\ }

\newcommand{\depend}[1]{\mathit{depend}(#1)}
\newcommand{\source}[1]{\mathit{source}(#1)}
\newcommand{\target}[1]{\mathit{target}(#1)}

\begin{document}

\title{Precisely Analyzing Loss in Interface Adapter Chains}
\author{Yoo Chung}
\maketitle

\begin{abstract}
  Interface adaptation allows code written for one interface to be used with a software component with another interface.  When multiple adapters are chained together to make certain adaptations possible, we need a way to analyze how well the adaptation is done in case there are more than one chains that can be used.  We introduce an approach to precisely analyzing the loss in an interface adapter chain using a simple form of abstract interpretation.
\end{abstract}

\section{Introduction}
\label{sec:introduction}

The computing world of today is becoming increasingly diverse.  This includes a great increase in the number of services that are being developed that provide various functionality.  It also includes an increase in the number of developers independently creating new interfaces to similar computing services.  To avoid lock-in to a specific vendor, it would be ideal if code written to the interface of one service can be used without change when it needs to be used with another service.  Unfortunately, this cannot be done directly without standardization of interfaces, and the slow speed of standardization processes coupled with the rapid proliferation of services precludes standardization in many cases.

Interface adapters can provide a solution to this problem by transforming interfaces as necessary.  And chaining them together enables much more flexibility without incurring a prohibitive development cost in creating all of the required interface adapters for direct interface adaptation.  Unfortunately, an interface adapter is likely to be imperfect, so interface adaptation would often incur adaptation loss.  This is even more of an issue when interface adapters are chained.  To properly consider loss in the construction of interface adapter chains, a mathematical framework is required to analyze such chains.

Previous work either took the approach of treating a method as a single unit that is either available or not available, or loosened this by treating the availability of a method probabilistically~\cite{chung:ietsoftw2010,chung:icsess2010}.  The former is unable to precisely analyze cases where individual methods cannot be precisely adapted, whereas the latter is a very rough approximation.  This paper outlines another approach to handling the partial adaptation of methods, which is to use a simple form of abstract interpretation~\cite{h:cousot:popl1977,jones1995-abstract-interp,cousot:csur1996,dams1997-toplas}.  Dividing arguments up into abstract domains, the abstract interpretation approach looks at which abstract domains can be handled by a method instead of looking at the likelihood.  Interface adapters are ``executed abstractly'' to derive which abstract domains can be handled by a method in the target interface.

The paper is organized as follows.  We look at related work in section~\ref{sec:related}.  In section~\ref{sec:abstract-math}, we describe the mathematical framework with which to analyze loss in interface adapter chains using abstract interpretation.  We discuss the computational complexity of the problem in section~\ref{sec:complexity-abstract} and describe a greedy algorithm in section~\ref{sec:greedy-abstract-algorithm}.  Section~\ref{sec:conclusions} concludes.

\section{Related work}
\label{sec:related}

Vayssi\'{e}re~\cite{vayssiere:doa2001} supports the interface adaptation of proxy objects for Jini~\cite{jini}.  The goal is to enable clients to use services even when they have different interfaces than expected.  Adapters are registered with an adapter service, which in turn registers all acyclic chains of interface adapters that can adapt to a given type to the lookup service.  Registering all possible chains can result in an exponential number of registrations in the lookup service (they claim that the maximum number of registrations would be limited, but they ignore that there can be an exponential number of paths in a reasonable graph).  There is no discussion of which chain of adapters should be used to adapt an interface, simply specifying that all chains matching the expected input and output types should be returned.

Gschwind~\cite{gschwind:sem2002} allows components to be accessed through a foreign interface and implements an interface adaptation system for Enterprise JavaBeans~\cite{ejb}.  It implements a centralized adapter repository that stores adapters, along with weights that mark the priority of an adapter.  Clients query an adaptation component to obtain an interface adapter chain, which is used to convert the interface of an object into another.  The adaptation repository uses Dijkstra's algorithm~\cite{dijkstra:mathematik1959} to construct the shortest interface adapter chain that adapts a source interface into a target interface.  While there is support for marking an adapter as lossy or not, it does not have the capability to properly analyze and compare the loss of interface adapter chains.

Ponnekanti and Fox~\cite{ponnekanti:percom2003} suggests using interface adapter chaining for network services to handle the different interfaces available for similar types of services.  They provide a way to query all services whose interfaces can be adapted to a known interface.  They also support lossy adapters, but the support is limited to detecting whether a particular method and specific parameters can be handled at runtime.  They do not provide a method to analyze the loss of an interface adapter chain, so they are unable to choose a chain with less loss when alternatives are available.  Adapter chains are constructed through a rule system based on the source and target interfaces of adapters, which is similar to constructing a path in a graph through a blind search algorithm.  They also support the composition of services in addition to transforming a single interface to another, which is accomplished by constructing a tree of interface adapters.

Kim et al.~\cite{kim:percom2008} describes an ad~hoc scheme for analyzing the loss in interface adapter chains.  The scheme is based on boolean matrixes which specify the methods required in a source interface to implement a method in a target interface.  A mapping product is defined on these matrixes which computes the loss incurred when interface adapters are chained.  The mathematical model they use is not rigorously constructed, however.  They also only consider the adaptation of methods as a whole and do not handle the case where methods could handle certain arguments but not others.

\section{Mathematical basics}
\label{sec:abstract-math}

We can define an interface adapter graph, which is a directed graph where interfaces are nodes and adapters are edges.  We also assume that a method accepts only a single argument: multiple arguments can be modeled as a tuple with multiple components~\cite{multiple-arguments-as-tuple}, while no argument can be modeled by a dummy argument.

For each method in an interface, its argument domain is divided up into disjoint domains that are each represented by abstract values.  For example, integer arguments could be mapped to abstract values $d_+$, $d_-$, and $d_0$.  These domains and abstract values must be fixed for each method in an interface, and they must not be different for different interface adapters.  The chaining of interface adapters cannot be analyzed otherwise.

There is a special abstract value denoted by $\bot$ distinct from any other abstract value, which is used when a method is unable to handle any other abstract value, \ie when a method cannot handle any arguments.  Including $\bot$ in the set of all possible abstract values for a method can be considered lifting the set~\cite{lifted-set}, and we call this lifted set the \emph{abstract argument domain} for the method.

\subsection{Method dependencies}
\label{sec:abstract-dependencies}

To represent how methods in a source interface are used to implement a method in a target interface, including which arguments must be used for the methods in the source interface to handle an argument for the method in the target interface, we use a function-based approach.  Functions are represented with a set-theoretic approach~\cite{set-function}, where a function is a set of pairs of arguments and values.

Let there be an interface adapter~$A$ with source interface~$I_S$ and target interface~$I_T$.  Let $D_i$ be the abstract argument domain of method~$i$ in $I_S$, and let $D'_j$ be the abstract argument domain of method~$j$ in $I_T$.  If $I_S$ has $n$~methods and $I_T$ has $n'$~methods, then we can define the \emph{abstract dependency function}~$h_A$ for $A$:
\[ h_A : D_1 \times D_2 \times \cdots \times D_n \rightarrow 2^{D'_1} \times 2^{D'_2} \times \cdots \times 2^{D'_{n'}} \]
where the $j$th component in a result tuple is the set of all possible abstract values that can be handled by method~$j$ in the target interface given the specified abstract values accepted by the methods in the source interface.  If a method is unavailable in the source interface or it is unable to handle any arguments, then its corresponding abstract value in the argument to $h_A$ would be $\bot$.  All components in the result tuple include $\bot$ so that chained interface adapters can be analyzed even when certain methods end up being unavailable.

We also define a method availability vector~$p$ for the abstract interpretation approach.  It is simply a tuple of sets with type $2^{D_1} \times 2^{D_2} \times \cdots \times 2^{D_{n}}$, assuming there are $n$~methods and $D_i$ is the abstract argument domain for method~$i$.  It represents the abstract values that each method is able to handle, and it is not intrinsic to an interface or interface adapter.  Instead, it expresses how well interface adaptation is done.

In the following, we will use the shorthand $\cup$ for the ``union'' of two tuples, in which the result is a tuple with the same number of components, and each component is the union of the corresponding components in the arguments.  In other words, $[S_1, \ldots, S_n] \cup [S'_1, \ldots, S'_n] = [S_1 \cup S'_1, \ldots, S_n \cup S'_n]$.  The ``indexed union'' of tuples is a straightforward extension of the shorthand.  We will also denote the Cartesian product of the components of a tuple of sets with the prefix~$\times$.  In other words, $\times[S_1, \ldots, S_n] = S_1 \times \cdots \times S_n$.

If an interface adapter~$A$ with abstract dependency function~$h_A$ were to be used to convert an source interface~$I_S$ to target interface~$I_T$, where the method availability vector for $I_S$ is $p$, then it is easy to see that each component of the resulting method availability vector~$q$ should be:
\begin{equation}
  \label{eq:abstract-dependency-adaptation}
  q = \bigcup_{x \in (\times p)} h_A(x)
\end{equation}
Basically, each component of $q$ should be the union of all possible abstract values for the corresponding method as adapted from $p$ through $h_A$.

It would be more convenient to simply apply a function to a method availability vector, and this can easily be done by defining an \emph{abstract adaptation function}~$f_A$ corresponding to an abstract dependency function~$h_A$ based on equation~(\ref{eq:abstract-dependency-adaptation}), where $D_1$, \ldots, $D_n$ are the abstract argument domains for the source interface of the adapter:
\begin{equation}
  \label{eq:abstract-adaptation-function}
  f_A = \{ (X, \bigcup_{x \in X} h_A(x)) \,|\, X \subseteq D_1 \times \cdots \times D_n \}
\end{equation}

Obviously, constructing an abstract dependency function for an interface adapter is easier than constructing an abstract adaptation function, so it would be expected that the abstract dependency function is constructed first, and then the abstract adaptation function is constructed from this, after which the abstract adaptation function would be used to analyze interface adapter chains.  We will denote the abstract adaptation function of an interface adapter~$A$ by $\depend{A}$.

Now that we have defined the abstract adaptation function, defining the adaptation operator is simple: it is simply function application.  In fact, we will not use special notation to represent the operator and will just use the standard notation:
\[ f_A(p) \]

Unlike in previous work which used a single value to denote the availability of a method~\cite{chung:ietsoftw2010,chung:icsess2010}, there is no need to define something like a dummy method since an adaptation dependency function is general enough to encompass cases which are ambiguous in the other approaches.  If a method in a target interface can always be implemented, then the appropriate abstract value results can be specified for an argument of $[\bot, \ldots, \bot]$, whereas if a method can never be implemented, then $[\{\bot\}, \ldots, \{\bot\}]$ can be returned as the result.

We denote a method availability vector for an interface~$I$ to a fully functional service by $\mathbf{1}_I$, where each set in the tuple includes the entire abstract argument domain for the corresponding method.

\subsection{Adapter composition}
\label{sec:abstract-composition}

We would like to be able to derive a composite abstract adaptation function from the composition of two abstract adaptation functions, which would be equivalent to describing the chaining of two interface adapters as if they were a single interface adapter.

For the abstract interpretation approach, this is very straightforward because the abstract adaptation function is a function: the composition is just function composition.  And function composition is well known to be associative~\cite{function-composition}, although not commutative in general. In fact, we will use the standard notation for abstract adaptation function composition:
\[ f_{A'} \circ f_A \]

We can also show a monotonicity property, which formalizes the notion that extending an interface adapter chain results in worse adaptation loss.  If $A_1$ and $A_2$ are interface adapters, where $A_1$ converts $I_1$ to $I_2$ and $A_2$ converts $I_2$ to $I_3$, with $f_{A_1} = \depend{A_1}$ and $f_{A_2} = \depend{A_2}$, and $h_{A_1}$ and $h_{A_2}$ are the adaptation dependency functions associated with $f_{A_1}$ and $f_{A_2}$, respectively, then:
\[ f_{A_2}(z) = \bigcup_{x \in (\times z)} h_{A_2}(x) \subseteq \bigcup_{x \in (\times \mathbf{1}_{I_2})} h_{A_2}(x) = f_{A_2}(\mathbf{1}_{I_2}) \]
\begin{equation}
  \label{eq:abstract-mononicity}
  \therefore \; (f_{A_2} \circ f_{A_1})(\mathbf{1}_{I_1}) \subseteq f_{A_2}(\mathbf{1}_{I_2})
\end{equation}
with the shorthand that the ``subset'' relationship for tuples denotes each corresponding component satisfying the subset relationship, \ie \( [S_1, \ldots, S_n] \subseteq [S'_1, \ldots, S'_n] \) denotes \( S_1 \subseteq S'_1 \wedge \cdots \wedge S_n \subseteq S'_n \).

The definitions of the abstract adaptation function, the abstract dependency function, and the method availability vector in section~\ref{sec:abstract-dependencies}, along with the associativity and monotonicity rules proven in this section, provide a succinct way to mathematically express the chaining of lossy interface adapters using an abstract interpretation approach.

\subsection{An example}
\label{sec:abstract-example}

\begin{figure}
  \centering
  \includegraphics[width=10cm]{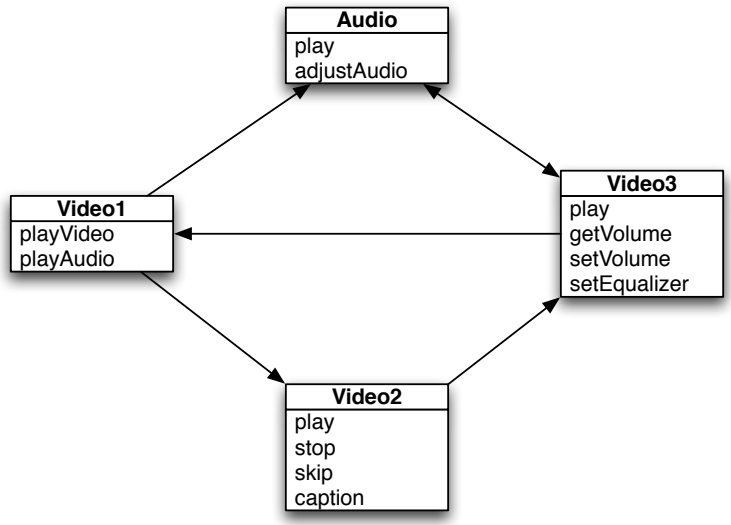}
  \caption{Interface adapter graph example.}
  \label{fig:multiple-interface-example}
\end{figure}

As an example, we apply the abstract interpretation approach to the interfaces and adapters in figure~\ref{fig:multiple-interface-example}.  The abstract argument domains for each method in each interface must be determined manually based on a human-level understading of the interfaces, not only considering a natural division of arguments but also anticipating divisions relevant to potential interface adapters.  Table~\ref{tab:abstract-example-domains} contains an example of how the abstract argument domains could be defined for figure~\ref{fig:multiple-interface-example}.

\begin{table}
 \centering
 \begin{tabular}{lll}
   Interface & Method & Abstract values \\ \hline
   \multirow{2}{*}{\textit{Video1}} & \textit{playVideo} & MOV, AVI, MKV \\
   & \textit{playAudio} & MP3, OGG, WAV \\ \hline
   \multirow{4}{*}{\textit{Video2}} & \textit{play} & INDEO, MP4, THEORA, DIVX \\
   & \textit{stop} & DUMMY \\
   & \textit{skip} & INTEGER \\
   & \textit{caption} & LANGUAGE \\ \hline
   \multirow{4}{*}{\textit{Video3}} & \textit{play} & MOV, AVI, MKV, RM \\
   & \textit{getVolume} & DUMMY \\
   & \textit{setVolume} & INTEGER \\
   & \textit{setEqualizer} & EQSPEC \\ \hline
   \multirow{2}{*}{\textit{Audio}} & \textit{play} & AU, WAV, OGG \\
   & \textit{adjustAudio} & VOLUME, EQUALIZER, MIXED
 \end{tabular}
 \caption{Example abstract argument domains for figure~\ref{fig:multiple-interface-example}.}
 \label{tab:abstract-example-domains}
\end{table}

For interface adapter \textit{Video1toVideo2}, the \textit{playAudio} method in \textit{Video1} is not required to implement any of the methods, while \textit{stop}, \textit{skip}, and \textit{caption} methods in \textit{Video2} cannot be implemented using the methods of \textit{Video1}.  Let us say that \textit{playVideo} of \textit{Video1} can handle MOV files, which can contain video encoded in MP4, AVI files, which can contain video encoded in INDEO and DIVX, and MKV files, which can contain video encoded in MP4, DIVX, and THEORA.  Then we would have an abstract dependency function with tuples as specified in table~\ref{tab:abstract-example-dependency}.  The abstract dependency function has 16~elements, while the abstract adaptation function has $2^4 \times 2^4 = 256$ elements, which we do not show here.

\begin{table}
  \newcommand{\ABmov}{\mathrm{MOV}}
  \newcommand{\ABavi}{\mathrm{AVI}}
  \newcommand{\ABmkv}{\mathrm{MKV}}
  \newcommand{\ABrm}{\mathrm{RM}}
  \newcommand{\ABmpt}{\mathrm{MP3}}
  \newcommand{\ABogg}{\mathrm{OGG}}
  \newcommand{\ABwav}{\mathrm{WAV}}
  \newcommand{\ABau}{\mathrm{AU}}
  \newcommand{\ABindeo}{\mathrm{INDEO}}
  \newcommand{\ABmpf}{\mathrm{MP4}}
  \newcommand{\ABtheora}{\mathrm{THEORA}}
  \newcommand{\ABdivx}{\mathrm{DIVX}}
  \newcommand{\ABdummy}{\mathrm{DUMMY}}
  \newcommand{\ABinteger}{\mathrm{INTEGER}}
  \newcommand{\ABeqspec}{\mathrm{EQSPEC}}
  \newcommand{\ABvolume}{\mathrm{VOLUME}}
  \newcommand{\ABequalizer}{\mathrm{EQUALIZER}}
  \newcommand{\ABmixed}{\mathrm{MIXED}}

  \centering
  \begin{displaymath}
    \begin{array}{c}
      ((\bot, \bot), (\{\bot\}, \{\bot\}, \{\bot\}, \{\bot\})) \\
      ((\bot, \ABmpt), (\{\bot\}, \{\bot\}, \{\bot\}, \{\bot\})) \\
      ((\bot, \ABogg, (\{\bot\}, \{\bot\}, \{\bot\}, \{\bot\})) \\
      ((\bot, \ABwav), (\{\bot\}, \{\bot\}, \{\bot\}, \{\bot\})) \\
      ((\ABmov, \bot), (\{ \bot, \ABmpf \}, \{\bot\}, \{\bot\}, \{\bot\})) \\
      ((\ABmov, \ABmpt), (\{ \bot, \ABmpf \}, \{\bot\}, \{\bot\}, \{\bot\})) \\
      ((\ABmov, \ABogg), (\{ \bot, \ABmpf \}, \{\bot\}, \{\bot\}, \{\bot\})) \\
      ((\ABmov, \ABwav), (\{ \bot, \ABmpf \}, \{\bot\}, \{\bot\}, \{\bot\})) \\
      ((\ABavi, \bot), (\{ \bot, \ABindeo, \ABdivx \}, \{\bot\}, \{\bot\}, \{\bot\})) \\
      ((\ABavi, \ABmpt), (\{ \bot, \ABindeo, \ABdivx \}, \{\bot\}, \{\bot\}, \{\bot\})) \\
      ((\ABavi, \ABogg), (\{ \bot, \ABindeo, \ABdivx \}, \{\bot\}, \{\bot\}, \{\bot\})) \\
      ((\ABavi, \ABwav), (\{ \bot, \ABindeo, \ABdivx \}, \{\bot\}, \{\bot\}, \{\bot\})) \\
      ((\ABmkv, \bot), (\{ \bot, \ABmpf, \ABdivx, \ABtheora \}, \{\bot\}, \{\bot\}, \{\bot\})) \\
      ((\ABmkv, \ABmpt), (\{ \bot, \ABmpf, \ABdivx, \ABtheora \}, \{\bot\}, \{\bot\}, \{\bot\})) \\
      ((\ABmkv, \ABogg), (\{ \bot, \ABmpf, \ABdivx, \ABtheora \}, \{\bot\}, \{\bot\}, \{\bot\})) \\
      ((\ABmkv, \ABwav), (\{ \bot, \ABmpf, \ABdivx, \ABtheora \}, \{\bot\}, \{\bot\}, \{\bot\}))
    \end{array}
  \end{displaymath}
  \caption{Elements in an example abstract dependency function.}
  \label{tab:abstract-example-dependency}
\end{table}

From the adaptation dependency function in table~\ref{tab:abstract-example-dependency}, it is easy to see that with the interface adapter \textit{Video1\-to\-Video2}, an argument for \textit{play} in \textit{Video2} that corresponds to abstract value MP4 can be handled if \textit{playVideo} in \textit{Video1} can handle an argument that corresponds to abstract values MOV or MKV.  If $f$ is the corresponding abstract adaptation function and the method availability vector is $p = [\{ \bot, \mathrm{MOV}, \mathrm{MKV} \}, \{\bot, \mathrm{MP3}\}]$, then 
\[ f(p) = [ \{ \bot, \mathrm{MP4}, \mathrm{DIVX}, \mathrm{THEORA} \}, \{\bot\}, \{\bot\}, \{\bot\} ] \]
which shows that with interface adaptation to \textit{Video2}, \textit{play} would be able to handle arguments corresponding to MP4, DIVX, and THEORA, while the \textit{stop}, \textit{skip}, and \textit{caption} methods would not be available.

For the other interface adapters, the adapters \textit{Video2\-to\-Video3}, \textit{Video3\-to\-Audio}, and \textit{Video3\-to\-Video1} have 40~elements in their abstract dependency functions and 2048~elements in their abstract adaptation functions, while the adapters \textit{Video1\-to\-Audio} and \textit{Audio\-to\-Video3} have 16~elements in their abstract dependency functions and 256~elements in their abstract adaptation functions.

\section{Complexity}
\label{sec:complexity-abstract}

The abstract interpretation approach can easily encompass the discrete approach~\cite{chung:ietsoftw2010} by having each method accept only a single abstract value besides $\bot$.  However, this does not mean that the optimal adapter chaining problem, where the number of accepted abstract values is maximized, is NP-complete.  The reason is that the reduction of a method dependency matrix in the discrete approach to an abstract adaptation function in the abstract interpretation approach can require exponential time.  In fact, simply storing the abstract adaptation function can require an exponential amount of memory.

In the discrete and probabilistic approaches~\cite{chung:ietsoftw2010,chung:icsess2010}, representing a method dependency matrix or a probabilistic adaptation factor requires $O(m^2)$~space, where $m$ is the maximum number of methods in an interface.  Applying the adaptation or composition operators would require $O(m^2)$ or $O(m^3)$~time.  However, since the abstract interpretation approach requires that functions be represented as a set of tuples to retain generality, it requires $O(d^m)$~space to represent an abstract dependency function and $O(2^{dm})$~space to represent an abstract adaptation function.  Recall that an abstract argument domain must include $\bot$ in addition to a separate abstract value, so $d$ is necessarily greater than or equal to 2, thus this is a truly exponential bound.

In fact, the exponential space complexity is a lower bound, not just an upper bound.  With exactly $m$~methods in a source interface and with $d_1$, $d_2$, \ldots, $d_m$ abstract values in the abstract argument domains for each method, the number of tuples in the abstract dependency function is exactly $\prod_{i=1}^m d_i$, and the number of tuples in the abstract adaptation function is exactly $\prod_{i=1}^m 2^{d_i}$.  We can see this in the example of section~\ref{sec:abstract-example}.

With complex interfaces that have many methods and non-trivial abstract argument domains, the exponential space complexity of the abstract interpretation approach makes it unlikely to be used in a system not dedicated to analyzing lossy interface chains which lacks the correspondingly exponential amount of memory.  It is an open question whether real world interfaces will be trivial enough such that the abstract interpretation approach can be used more generally.

\section{A greedy algorithm}
\label{sec:greedy-abstract-algorithm}

While the exponential complexity of the abstract interpretation approach might make it unfeasible to obtain an optimal adapter chain on demand in a resource-constrained interactive system and virtually impossible to analyze complex interface adapters, it may be a slow but useful tool for software architecture analysis to derive an optimal adapter chain with simple interface adapters.  So we describe an algorithm which can construct an optimal interface adapter chain in algorithm~\ref{alg:abstract-greedy-algorithm}, which works thanks to the monotonicity property in equation~(\ref{eq:abstract-mononicity}).

\begin{algorithm}
  \caption{Adapter chaining algorithm with abstract interpretation.}
  \label{alg:abstract-greedy-algorithm}
  \begin{algorithmic}
  \Procedure{Prob-Greedy-Chain}{$G = (V, E)$, $s$, $t$}
    \State $\mathit{C} \gets \{ [] \}$
    \Comment{chains to extend}
    \State $\mathit{M} = \emptyset$
    \Comment{discarded chains}
    \State $D \gets \{[] \mapsto \mathbf{I}_{\dim(\mathbf{1}_T)} \}$
    \Comment{method dependency matrixes}
    \While{$C \neq \emptyset$}
      \State $c \gets \mbox{element of $C$ maximizing $\textsc{Count-Abstract}(c,D)$}$
      \If{$c \neq [] \wedge \source{c[1]} = s$}
        \State \textbf{return} $c$
      \ElsIf{no acyclic chain not in $C \cup M$ extends $c$}
        \State $C \gets C - \{c\}$
        \State $M \gets M \cup \{c\}$
      \Else
        \If{$c = []$}
          \State $B \gets \{ [e] \,|\, e \in E, \target{e} = t\}$
        \Else
          \State $B \gets \{ e : c \,|\, e \in E, \target{e} = \source{c[1]} \}$
        \EndIf
        \State remove cyclic chains from $B$
        \State $C \gets C \cup B$
        \State $D \gets D \cup \{ e:c \mapsto D[c] \circ \depend{e} \,|\, e : c \in B \}$
      \EndIf
    \EndWhile
  \EndProcedure
  \end{algorithmic}
\end{algorithm}

\begin{algorithm}
  \caption{Computing number of accepted abstract values.}
  \label{alg:abstract-count}
  \begin{algorithmic}
  \Function{Count-Abstract}{$c$, $D$}
    \State $s \gets \source{c[1]}$
    \State $v \gets D[c](\mathbf{1}_s)$
    \State \textbf{return} sum of count of abstract values in each component of $v$
  \EndFunction
  \end{algorithmic}
\end{algorithm}

Algorithm~\ref{alg:abstract-greedy-algorithm} can be extended to support behavior similar to service discovery by checking whether the current source is among a potential set of source interfaces instead of just checking against one can be similarly extended.  Abstract values that represent arguments can also be similarly weighted to prioritize what needs to be adapted.

\section{Conclusions}
\label{sec:conclusions}

When many similar software components from different developers need to be used interchangeably due to various reasons such as having to work with diverse software environments or an evolving software architecture, using interface adapters is a way to reuse existing software components without having to rewrite them.  Chaining interface adapters allows much more interface adaptation with fewer resources having to be devoted to actually developing interface adapters.  However, insurmountable incompatibilities between interfaces of similar software components may force interface adapters to be less than perfect, so an approach to analyze the loss incurred by chains of interface adapters is needed.

The abstract interpretation approach can be the most precise way to analyze the loss in chains of interface adapters, not having to rely on questionable assumptions about the partial adaptation of methods.  However, it does rely on abstract argument domains being properly defined for every method in every interface.  An improperly set up abstract argument domain would require a great deal of effort to update related data structures appropriately, and sometimes it might not be easy to define an abstract argument domain which would work well with every conceivable interface and interface adapter.

A much more serious problem with the abstract interpretation approach is its complexity.  It has exponential space complexity, which means a prohibitive amount of memory may be required.  Even worse, it requires exponential effort to set up the necessary values, which for even moderately complex interface adapters might be infeasible to do automatically by computer, much less by a human developer.  And the necessary functions have to be constructed by a human developer unless sophisticated program analyses can be developed that could automatically define abstract argument domains and infer how interface adapter code will adapt them.

The exponential complexity of the abstract interpretation approach suggests that it should be used for offline analysis of simple interface adapter chains.  The better precision may be useful when determining if a set of interface adapters shipped with a deployment of a ubiquitous computing environment can satisfactorily support seamless operation.  However, the abstract interpretation approach may still be practical as a subsystem if the number of methods and the size of abstract argument domains are small: it is an open question of whether they would be small enough in real world systems.

These limitations suggest future avenues for research.  One direction would be to take advantage of the semantics of a software component to automate the setup of abstract argument domains as much as possible.  Another is the study of alternative formulations for expressing abstract argument domains besides sets or specialized adaptation functions applicable to a wide range of software interfaces, which could allow polynomial complexity for analyzing loss in interface adapter chains using abstract interpretation.

\bibliographystyle{plain}
\bibliography{strings,articles,hearsay,local,proceedings,books}

\end{document}